\documentclass[prd,nofootinbib,12pt]{revtex4}

\usepackage{lineno,hyperref}
\usepackage{amsfonts,amssymb,amsmath}
\usepackage{epsfig}
\usepackage{mathrsfs}
\usepackage{amssymb}
\usepackage{graphicx}
\usepackage{amsmath}
\usepackage{graphicx}
\usepackage{amsmath}
\usepackage{xcolor}
\usepackage{color}

\begin{document}

\title{The primordial black hole from running curvaton }

\author{Lei-Hua Liu$^{1}$}
\email{liuleihua8899@hotmail.com}

\affiliation{$^1$Department of Physics, College of Physics, Mechanical and Electrical Engineering, Jishou University, Jishou 416000, China}

\begin{abstract}

	In light of our previous work \cite{Liu:2019xhn}, we investigate the possibility of the formation of a primordial black hole in the second inflationary process induced by the oscillation of curvaton. By adopting the instability of the Mathieu equation, one could utilize the $\delta$ function to fully describe the power spectrum. Due to the running of curvaton mass, we can simulate the value of abundance of primordial black holes nearly covering all of the mass ranges, in which we have given three special cases. One case could account for the dark matter in some sense since the abundance of a primordial black hole is about $75\%$. At late times, the relic of exponential potential could be approximated to a constant of the order of cosmological constant dubbed as a role of dark energy. Thus, our model could unify dark energy and dark matter from the perspective of phenomenology.  Finally, it sheds new light on exploring Higgs physics.

\end{abstract}

\maketitle


\section{Introduction}	

The formation of a primordial black hole (PBH) in an inflationary period consisting of preheating period offers an intriguing window for exploring the early universe \cite{Hawking:1971ei,Carr:1974nx,Grindlay:1975eb}. From another perspective, the PBH could account for the formation of dark matter (DM) according to the abundance of PBH \cite{Ivanov:1994pa,Carr:2016drx,Gaggero:2016dpq,Inomata:2017okj,Kovetz:2017rvv,Georg:2017mqk}. The explicit observation of gravitational waves is the most essential achievement by the joint LIGO/Virgo collaboration, especially for the merge of black holes (BHs) \cite{LIGOScientific:2016aoc,LIGOScientific:2016emj,LIGOScientific:2017bnn}, in which there are some merged BHs whose mass range is around $30 M_{\odot}$ that cannot be achieved by the stellar evolution. However, the mass range of PBH could cover this mass, thus the PBH is also of significance to exploring the formation of BHs \cite{Bird:2016dcv,Sasaki:2016jop,Carr:2016drx}. The range of mass from $10^{-17}-10^{-15} M_{\odot}$ and $10^{-15}-10^{-13} M_{\odot}$ could almost explain the origin of DM since these scales cannot be constrained by the observations.  PBH is usually formed due to the gravitational collapse of the overdense region. As a consequence, it will lead to the large amplitude of curvature perturbation at some certain scales, which can be realized by tuning the background dynamics of quantum field in inflationary universe \cite{Garcia-Bellido:1996mdl,Garcia-Bellido:2017mdw,Domcke:2017fix,Kannike:2017bxn,Carr:2017edp,Ballesteros:2017fsr,Hertzberg:2017dkh,Franciolini:2018vbk,Kohri:2018qtx,Ozsoy:2018flq,Biagetti:2018pjj}. Due to its specification, PBH could leave the imprints for the observation, $\it e.g.$ the seeds for the formation of galaxy, the evaporation of PBH could interestingly interpret the observations of point-like gamma-ray sources \cite{Belotsky:2014kca,Khlopov:2008qy}.

There are several ways for realizing the mechanism of enhancing the power spectrum at small scales. One effective way named by ultra-slow-roll inflation \cite{Martin:2012pe,Motohashi:2014ppa}, especially for the inflection point of the potential \cite{Germani:2017bcs} dubbed as a very economical way of achieving the ultra-slow-roll inflation for generating the enhanced power spectrum \cite{Motohashi:2017kbs,Ezquiaga:2017fvi,Ballesteros:2018wlw}. Under this framework, it is not explicitly realized a viable mechanism for keeping the e-folding number is around $50-60$ \cite{Passaglia:2018ixg,Sasaki:2018dmp}. Another way of enhancing the power spectrum is implemented by the non-minimal coupling and noncanonical kinetic term \cite{Fu:2019ttf,Fu:2019vqc,Dalianis:2019vit,Lin:2020goi,Braglia:2020eai,Gundhi:2020zvb,Cheong:2019vzl}, in which the potential and the non-minimally coupling function should have some special form with fine-tuning parameters of various models. In light of $k$ inflation \cite{Armendariz-Picon:1999hyi,Garriga:1999vw} and $G$ inflation \cite{Kobayashi:2010cm,Kobayashi:2011nu}. In particular, there is a new mechanism for generating the large power spectrum under the framework of noncanonical kinetic terms at small scales is called the resonant sound speed during inflation, being different from the standard procedure for the preheating period, in which its new place is the resonant sound speed appears during the early inflationary period ($\Delta N$ is around $10$) \cite{Cai:2018tuh}. Due to its simplicity and richness from the phenomenology, it can be applied to the stochastic gravitational waves \cite{Cai:2019jah} and this mechanism can be embedded into DBI inflation \cite{Chen:2020uhe} and curvaton-inflaton mixed inflation  \cite{Chen:2019zza}. A similar mechanism was proposed for which the sound speed approaches zero at some stage of inflation in single field inflation \cite{Ballesteros:2018wlw,Kamenshchik:2018sig}. The amplification of curvature perturbation could also arise from the oscillation potential  \cite{Cai:2019bmk}. Even the PBH can be formed due to the peak theory \cite{Wang:2021kbh}.

To investigate the formation of PBH, we need the amplification of curvature perturbation resulting from the inflationary perturbation, in which these broad kinds of single-field inflationary theory are highly relevant to the shape of inflation. In order to relax this stringent condition, such kind of curvaton model was proposed, in which the curvature perturbation has arisen from the curvaton~\cite{Enqvist:2001zp, Moroi:2001ct, Lyth:2001nq} and it usually is considered as an independent field. Consequently, it could account for the various particle, $\it i.e. $ axion field \cite{Gong:2016yyb}, further it could from the axionic PBH due to the role of axion \cite{Kawasaki:2012wr,Ando:2018nge,Chen:2019zza}. In light of preheating process, we consider that the curvaton field comes via inflaton decay \cite{Traschen:1990sw,Kofman:1994rk,Shtanov:1994ce,Prokopec:1996rr,Greene:1997ge,Kofman:1997yn,Greene:1997fu}. If considering the multifield framework for curvaton \cite{Liu:2020zzv}, we could also investigate the impact from the noncanonical kinetic term. 
Moreover, our paper also aims to unify the DM and dark energy (DE) in light of \cite{Liu:2019xhn}. From a thorough investigation of the curvaton field, it will experience from its generation up to the DE epoch, we will give a full analysis of the curvaton evolution.

This paper is organized as follows. In section~\ref{revised model} we revise the running curvaton mechanism.  In section \ref{formation of pbh}, we will investigate the formation of PBH in light of the Mathieu equation. In section \ref{late time}, we give a simple analysis of the late time evolution of curvaton. Finally, section \ref{conclusion} gives the conclusion and discussions.

We work in natural units in which $c=1=\hbar$, but retain the Newton constant $G$.


\section{The revisit  of running curvaton model}
\label{revised model}
According to Ref. \cite{Liu:2019xhn}, we investigate the formation of PBH during preheating period.  The main feature for realizing this process is that there is a coupling between the curvaton field and the potential of inflation. Subsequently, one can utilize the  Mathieu equation to explore the formation of PBH during the preheating period, recently a similar mechanism for realizing the formation of PBH has been done in light of the stability of Mathieu equation \cite{Cai:2021wzd}.

\subsection{Action}
\label{action}
Before discussing the formation of PBH, we first revisit the running curvaton model. At the very beginning, there is only one field called inflaton, then the curvaton field will appear via the decay of inflaton. Meanwhile, the explicit coupling is between the inflationary potential and curvaton field being different from traditional coupling (Yukawa coupling for scalar fields). Moreover, we add an exponential potential for the curvaton field to depict the dynamical behavior of dark energy. Therefore, a total action can be written by 
\begin{equation}
S=\int d^{4}x\sqrt{-g}\bigg\{\frac{M_{P}^{2}}{2}R-\frac{1}{2}g^{\mu\nu}\nabla_{\text{\ensuremath{\mu}}}\phi\nabla_{\nu}\phi-\frac{1}{2}g^{\mu\nu}\nabla_{\text{\ensuremath{\mu}}}\chi\nabla_{\nu}\chi
-V(\phi)-\frac{g_0}{M_{P}^{2}}\chi^{2}V(\phi)-\lambda_{0}\exp[-\lambda_{1}\frac{\chi}{M_{P}}]\bigg\}
\label{total action}
\end{equation}
where $\chi$ and $\phi$ are curvaton and inflaton, respectively, $R$ presents the Ricci scalar and $g$ is the determinant of $g_{\mu\nu}$. and $g_0$, $\lambda_0$, and $\lambda_1$ are the dimensionless parameters determined by observations in Lagrangian, one point should be emphasized that $\lambda_0$ is implemented to mimic the dynamical behavior of dark energy, thus it is of order $10^{-120}$ in Planck units.
 And our curvaton model is realized in preheating period for which the inflaton will oscillate its potential, then the energy can be transformed into the curvaton field. As \cite{Carrion:2021yeh} showed, as the inflaton oscillates near its minimal point, it will lead to that the various inflationary potentials can be approximated by 
\begin{equation}
	V(\phi)\approx \frac{1}{2}m^2\phi^2,
	\label{approximating potential}
\end{equation}
where $m$ is the effective mass of inflaton determined by $\frac{d^2V(\phi)}{d\phi^2}$.  
Thus, our curvaton mechanism can be embedded into many inflationary models, in which action \eqref{total action} clearly shows that the explicit coupling term $\frac{g_0\chi^2}{M_{P}^2}V(\phi)$ is consisting of many forms. Due to this explicit coupling, the effective mass of curvaton can be easily obtained by 
\begin{equation}
m_{\rm \chi}^ 2\approx\frac{g_0}{M_{P}^2}V(\phi)
	\label{eff of chi}
\end{equation}
where we have neglected the contribution of the exponential part of curvaton's potential since it is too tiny before the inflaton finishes the decay. This mass parameter will play a significant role in simulating the formation of PBH in preheating period.

As we know, the wavelength is expanding slowly compared to the Hubble radius, thus the perturbation mode of curvaton will cease at the subhorizon. To re-enter the horizon, the second inflationary process is necessary for fulfilling the re-entering of the horizon.  In the next subsection, we will illustrate this process via the interaction between the curvaton and other particles. 

\subsection{The second inflation}
\label{second inflation}
The second inflation is highly relevant with the evolution of background (for inflaton and curvaton). In order to obtain a viable process of enhancing the power spectrum, we first need the interaction that is similar to action \eqref{total action}, 
\begin{equation}
	\mathcal{L}_{\rm int}=\frac{1}{2}m_{\chi}^2\chi^2+\frac{g_1}{M_{\rm Pl}}\varphi^2V(\chi),
	\label{interaction term}
\end{equation}
where $\varphi$ is one kind of scalar particles ($\it e.g.$ ultral light scalar particle) and $g_1$ is the dimensionless paramteter. 

Following the \cite{Moroi:2005np}, we will divide the preheating period into two parts: (a) The first preheating comes via the oscillation of inflaton after inflation which is named by $\phi$D or RD1 era. (b) The second inflation is driven by the curvaton potential energy after $\chi$D dubbed as $\chi$D or RD2 era. The condition for the occurrence of the second condition is that the initial amplitude of curvaton is large enough. The second thing we need to figure out for which background will be alternated by the second inflation process. As \cite{Moroi:2005np} pointed out, the curvaton model can be classified into three types: the large field type ($m_\chi> M_{\rm Pl}$), the small field type ($m_\chi< M_{\rm Pl}$) and the hybrid field type (the coupling between curvaton and inflaton). In some sense, they worked out the most general cases of curvaton scenario. In this paper, we focus on the hybrid field type. According to their explicit calculation, we could obtain that $N_e$ (e-folding number) will be enlarged to $70$ more or less, which is very important for our later analysis of the formation scale of PBH during the second inflation process. As for its observables, it will decrease the spectral index $n_\chi$ in the hybrid type of curvaton model. Here, we will focus on its curvature perturbation in the second inflation. 

After introducing the second inflation, the curvature perturbation induced by the curvaton will re-enter the Hubble radius. Consequently, the enhanced density perturbation could have longer time for collapsing into the PBH after re-enter the horizon.

\subsection{Instability of Mathieu equation}
\label{instability1}
In this subsection, we will obtain the EOM of $\delta\varphi$ field (quantum fluctuation of some spectator field) from action (\ref{total action}). Here, we define that $\varphi(x,t)=\bar{\varphi}(t)+\delta\varphi(x,t)$ where $\bar{\varphi}(t)$ denotes the background of spectator scalar particle only depending on time and $\delta\varphi(x,t)$ is the quantum fluctuations. After this definition, varying with action (\ref{total action}) and transferring into momentum space, one could obtain 
\begin{equation}
(\delta\ddot{\varphi}_{k}+3\frac{\dot{a}}{a}\delta\dot{\varphi}_{k})+\frac{k^{2}}{a^{2}}\delta\varphi_{k}+m^{2}_\chi\frac{g_1}{M_{P}^{2}}\chi^{2}\delta\varphi_{k}=0,
\label{eom of chi0}
\end{equation}
 there is an extra term $\frac{k^2}{a^2}\delta\varphi_k^2$ compared with EOM of $\bar{\varphi}(t)$, in which we will analyze the quantum fluctuations of the spectator field. 
 Subsequently, one could utilize the EOM of the background of curvaton field $\chi\approx \chi_0a^{-2/3}\sin(mt)$ in preheating period and rearrange of eq. (\ref{eom of chi0}). Then, we could obtain the following equation, 
\begin{equation}
\delta\ddot{\tilde{\varphi}}_{k}+(\frac{k^{2}}{a^{2}})\delta\tilde{\varphi}_{k}+\frac{g_0m^{2}_\chi}{2M_{P}^{2}}\chi_{0}^{2}\delta\tilde{\varphi}_k-\frac{m^{2}_\chi g_1}{2M_{P}^{2}}\delta\tilde{\varphi}_k\chi_{0}^{2}\cos(2m_{\chi}t)=0.
\label{eom of chi3 with cos0}
\end{equation}
 Finally, we set $z=m_{\rm \chi}t$ and $\delta\tilde{\varphi}_k=a^{3/2}\delta\varphi_k$, eq. (\ref{eom of chi3 with cos0}) becomes 
\begin{equation}
\delta\tilde{\varphi}_{k}''+(A_{k}-2q_{k}\cos[2z])\delta\tilde{\varphi}_{k}=0
\label{eom of background phi30}
\end{equation}
where the correspondence can be found as follows,
\begin{eqnarray}
A_{k}&=&\frac{k^{2}}{m_{\chi}^{2}a^{2}}+\frac{g_1\chi_{0}^{2}}{2M_{P}^{2}},\\
q_{k}&=&\frac{\chi_{0}^{2}g_1}{4M_{P}^{2}}.
\label{correspondence1}
\end{eqnarray} 
 Thus, eq. (\ref{correspondence1}) can become the standard form of the Mathieu equation. There is a key feature of Mathieu equation, in which the solution of eq. (\ref{eom of background phi30}) is proportional to 
\begin{equation}
\varphi_k\propto \exp[\mu_k^{(n)} z]=\exp[\mu_k^{(n)}m_{\rm \chi }t],
\label{instability}
\end{equation}
where $\mu_k^{(n)}$ is Lyapunov index depicting the $n$ th instability band of Mathieu equation, in which the first band is sufficient for investigating the production of PBH
and its corresponding formula at the first band is denoted by $\mu_k=\sqrt{\frac{q_k^2}{4}-(\frac{2k}{m_\chi}-1)^2}$ following the notation of Ref. \cite{Kofman:1997yn}. The resonance occurs at the $k=\frac{m_\chi}{2}$ and the first band could take the maximal value of $\mu_k=\frac{q_k}{2}$ considered in the following calculation (namely $m_\chi=2k$), where $m$ is the mass of inflaton in our paper, then solution $\chi_k$ will become
\begin{equation}
\varphi_k\propto \exp[\frac{q_k m_\chi t}{2}]=\exp[q_k kt],
\label{solution of chi}
\end{equation} 
which will be analyzed for the PBH formation, meanwhile we will observe that the formation of PBH is highly relevant with the curvaton mass $m_\chi$. Here, we have connected $k$ to curvaton mass $m$. Due to this instability (\ref{solution of chi}), it is a key for ensuring the formation of PBH at some certain scales which will be considered as an enhanced part of the power spectrum of curvaton, in which it will not impact the CMB observation \cite{Akrami:2018odb}.

In this section, we have revisited a large class of curvaton scenarios called running curvaton, in which there will be second inflationary process for forming the PBH as the initial amplitude of curvaton is large enough. During the second inflation, the mass of curvaton will play a significant role for the scale of PBH formation.

\section{Formation of PBH}
\label{formation of pbh}

In this section, we will proceed to the formation of PBH in light of the instability of the Mathieu equation. Before the detailed calculation, we first give a simple physical picture of the formation of PBH. Recall that the formation occurs during the preheating period, the energy of curvaton will be transferred into other spectator fields that derive the second inflation. The mass of curvaton will play a role in the instability band. 

To depict the super Hubble scale or the sub Hubble scale, it is determined by the comoving Hubble radius $\mathcal{H}=a H$ with $\mathcal{H}=\frac{da}{d\tau}$ ($\tau$ is the conformal time) and $H=\frac{da}{dt}$ ($t$ is the physical time). Combined with our analysis for the formation of PBH during the second inflation, the modes of quantum fluctuations of the curvaton field will re-enter the horizon ($\mathcal{H}^{-1}$ as the horizon). In our calculation, we utilize the physical time $t$. Thus, the evolution of power spectrum will be varying with respect to $\tilde{k}=\frac{k}{a}$ not $k$. 
The previous analysis has shown that the background of inflation will alternative, especially for which the e-folding number will be changed into $70$ more or less. Then, we will use this number as a reasonable input to analyze our Hubble radius.

\subsection{Power spectrum within the instability of Mathieu equation}
\label{power spectrum for instability}
One of the most essential ingredients for producing the PBH is enhancing the value of the power spectrum at certain scales beyond CMB constraints. In this paper, we will utilize the instability of the Mathieu equation to obtain the satisfying power spectrum.

To obtain the full power spectrum in light of this instability,  we should recall the content of the power spectrum utilizing $\delta N$ formalism\ \cite{Liu:2019xhn}, in which its corresponding formula is $P_\zeta=\frac{H_*^2}{9\pi^2}\frac{r_{\rm decay}^2}{\chi^2}$  ($H_*^2 $ denotes the value at freezing time for Hubble radius and here $\chi$ denotes the value of curvaton starts to oscillate), in which the range of $P_\zeta$ is consistent with observations as taking $0.12<r_{\rm decay}<1$. In a single field inflationary model, the power spectrum can be obtained by $P_\zeta=\frac{H^2_*}{8\pi^2\epsilon}$, in which $\epsilon$ is of order unity after inflation. Comparing these formulas for the power spectrum, it was easily concluded that they are consistent with each other once taking $0.12<r_{\rm decay}<1$. Consequently, we can utilize $P_\zeta=\frac{H^2_*}{8\pi^2\epsilon}$ as the first part of the full power spectrum, which could nicely recover the observations at large scales. Subsequently, we can follow the standard procedure to rewrite the $P_\zeta=\frac{H^2_*}{8\pi^2\epsilon}$
in terms of $A_s\big[\frac{\tilde{k}}{\tilde{k}_p}\big]^{n_s-1}$ in light of \cite{Baumann:2009ds}, where $\tilde{k}_p$ is a fiducial comoving momentum whose value is around $\tilde{k}_p=0.05~\rm Mpc^{-1}$.  

Here, we need to emphasize that the power spectrum of our model is related to the energy scale $k$, namely $k=\frac{m_{\chi}}{2}$. As for the observable power spectrum, it is highly relevant to the comoving momentum. Consequently, we need $k=a\tilde{k}$ relate the power spectrum to the observations.

From another perspective, we need the enhanced power                                                                                                                                                                        spectrum at some certain scales, to simulate the generation of PBH, similar mathematical property with \cite{Cai:2018tuh} (mechanism is different since their enhanced amplitude of curvature perturbation occurs at the inflation), the power spectrum (mainly from curvaton) will be experienced a exponential factor via instability (\ref{solution of chi}) at some certain scales $k_*$, $P_\zeta=\frac{H^2_*}{8\pi^2\epsilon}\exp[q_kmt]=\frac{H^2_*}{8\pi^2\epsilon}\exp[2q_kk_{\rm scale}t]=\frac{H^2_*}{8\pi^2\epsilon}\exp[2q_k a \tilde{k}_{\rm scale}t]$, where we have adopted the definition of $P_\zeta=k^3|\zeta_k|^2/(2\pi^2)$ and $|\zeta_k|\propto v_k$. Here, one can see that $k_{\rm scale}$ \footnote{Here, we adopt $k_{\rm scale}$ to distinguish the $k$ in eq. (\ref{the whole power spctrum}). } is some so-called certain scales explicitly related to the running mass $m$ via (\ref{instability}), the mass only needs to be smaller compared with the upper limits of inflationary potential from COBE normalization (analysis will be given later). Following \cite{Kofman:1997yn}, it clearly indicates that $\Delta k\propto q^l$ ($l$ is the l-th band of instability of Mathieu equation) and our case is $q\ll 1$.
 Consequently, the enhanced part of the power spectrum can be parametrized by the $\delta$ functions for some $k_{\rm scale}$.  Taking these two factors into account, one can get the full formula of the power spectrum as
\begin{eqnarray}
P_{\zeta}&=&A_{s}\bigg[\frac{k}{k_{p}}\bigg]^{n_{s}-1}\bigg[1+\frac{q_k}{2}\exp\big(2q_kk_{\rm scale}t\big)\delta(k-2k_{\rm scale})\bigg]\nonumber \\
&=&A_{s}\bigg[\frac{k}{k_{p}}\bigg]^{n_{s}-1}\bigg[1+\frac{q_{\rm k}}{2}\exp(q_{\rm k}mt)\delta(k-m_\chi)\bigg]\nonumber\\
&=&A_{s}\bigg[\frac{k}{k_{p}}\bigg]^{n_{s}-1}\bigg[1+\frac{q_{\rm k}}{2}\exp(q_{\rm k}mt)\delta(a\tilde{k}-m_\chi)\bigg],
\label{the whole power spctrum}
\end{eqnarray}
where $A_s =\frac{H^2_*}{8\pi^2\epsilon}$, $q_k$ is defined in eq. (\ref{correspondence1}) dubbed as the amplitude of enhanced power spectrum in some sense and $t$ is the physical time and the coefficient of an exponential factor of $\exp[q_kmt]$ comes via a triangle approximation. In the following calculation, we will use the running mass as an input manifesting the main feature of our model. Once obtaining this full power spectrum, one can numerically simulate its range within the observational constraints.

\begin{figure}[h!]
	\centering
	\includegraphics[height=9cm, width=10cm]{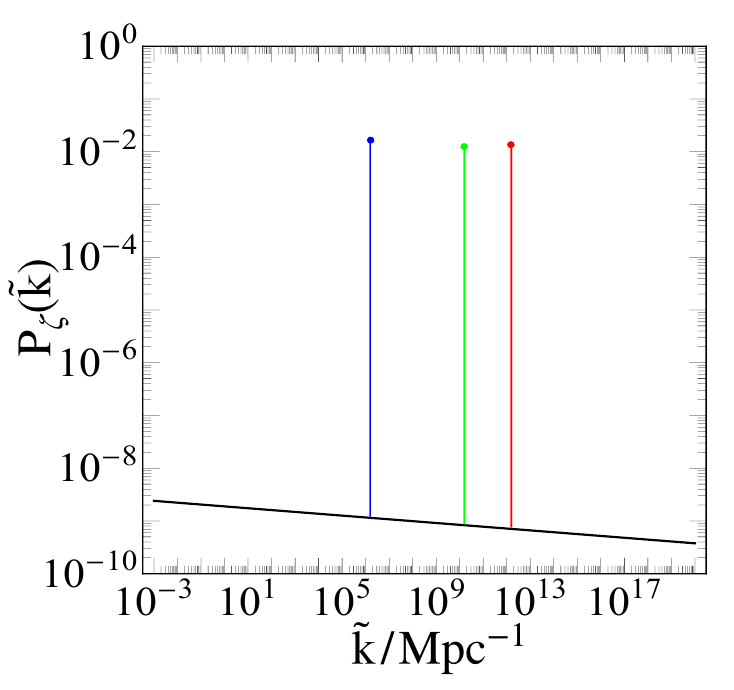}
	\vskip -0.4cm
	\caption{{\it The plot of power spectrum (\ref{the whole power spctrum}):} The horizontal line corresponds to energy scale whose range is $10^{-7} \rm~Mpc^{-1}\leqslant k\leqslant 10^{20}~\rm Mpc^{-1}$. The vertical line denotes the value of $P_\zeta$ whose range locates from $10^{-10}$ to $1$. The black solid line is observational constraints come via \cite{Akrami:2018odb}. The blue, green, red line correspond to the enhanced part of power spectrum as taking various mass of inflaton. $\tilde{k}_1=1.57\times 10^7~\rm Mpc^{-1}$ is for the blue point, $\tilde{k}_2=1.6\times 10^{10}~\rm Mpc^{-1}$ and $\tilde{k}_3=
		1.7\times 10^{12}~\rm Mpc^{-1}$ correspond to the green point, red point, respectively. The spectral index is set by $n_s=0.965$.  $q_k\ll 1$. }
	\label{full power spectrum}
\end{figure}

In figure \ref{full power spectrum}, it indicates that the varying trend of the full power spectrum (\ref{the whole power spctrum}), in which the black solid line corresponds to the observational value from COBE normalization \cite{Akrami:2018odb} whose order is of $10^{-9}$. In this figure, we give three values of $\tilde{k}$ as an illustration of forming PBH whose values are given in figure \ref{full power spectrum}. To relate the realistic energy scale, we unify the $\rm Kpc^{-1}$ into $\rm GeV$. We take $\tilde{k}_3$ as an example, in which it is straightforward for obtain $\tilde{k}_3\approx 10^{-26} \rm GeV$. Keeping in mind that our Universe has experienced exponential inflation and it is continuously expanding until the present, thus scale factor is a monotonically increasing function. Compared with reheating and preheating periods, the expansion of inflation is much larger. Thus, we could use $a(t_*)=\exp(70)$ (the inflation will last nearly $N=70$) as an approximation for obtaining the energy scale $k_*=a(t_*)\tilde{k}_3\approx 10^{13}~\rm GeV$ whose value is consistent with the energy scale of reheating or preheating period (depending on various models). The realistic value of $k_*$ will be larger than this value since we have set this approximation.

 As for the various value of $k$, it could be determined by the original definition according to $\mu_k=\sqrt{\frac{q_k^2}{4}-(\frac{2k}{m}-1)^2}$ as taking the maximal value for $\mu_k$. Especially for the enhanced part of the power spectrum, it reveals that its corresponding value could reach the order of $10^{-1}$ which is sufficient for the generation of PBH. All of these numerical simulations are done with $q_k\ll 1$ belonging to the narrow resonance. As for the choice of physical time, we set $t=10^{2}$ sec as a reasonable input since the formation of PBH occurs at the deep preheating period, in which it is after the matter-radiation equality with $10~\rm s<t<3~\rm min$.

Additionally,  this enhanced part of the power spectrum will not impact the observational constraints since the observational scale of CMB is approximately located from $10^{-5}~\rm Mpc^{-1}$ to $10~\rm Mpc^{-1}$, which leads to being consistent with observations.

\subsection{Formation of PBH in light of instability of Mathieu equation}
We will utilize the enhancement of the primordial power spectrum to investigate the formation of PBH in our theoretical framework. It can be seen clearly that we have large parameter spaces to construct the large amplitude of power spectrum to collapse into PBH, in which we give some specific values of curvaton mass and various $q_k$ whose value is much smaller than one. This formation process occurs during the radiation period induced by the oscillation of curvaton (the seoncond inflation). Meanwhile, the PBH will be generated after the perturbations of curvaton re-entry the horizon. The mass of PBH related to the horizon mass at the horizon re-entry with various co-moving wavenumber $k$ is 
\begin{equation}
	M(k)=\gamma \frac{4\pi}{\kappa^2 H}\approx M_{\odot}\bigg(\frac{\gamma}{0.2}\bigg)\bigg(\frac{g_*}{10.75}\bigg)^{-\frac{1}{6}}\bigg(\frac{k}{1.9\times 10^6~\rm Mpc^{-1}}\bigg),
	\label{mass of pbh}
\end{equation}
where $\kappa^{-1}=M_{\rm pl}=2.4\times 10^{18}~ \rm GeV$  is the reduced Planck mass, $H$ is evaluated at $k=a H$ ($H$ is Hubble parameter and $a$ is scale factor),  $\gamma$ is defined to the ratio of PBH mass to the horizon mass indicating the efficiency of collapse, in which its value is approximately set by $0.2$ from Ref. \cite{Carr:1975qj}, $g_*$ is the degrees freedom of energy densities at the formation of PBH. Since its formation occurs during the preheating period, 
the process happens during the deep radiation period whose value can be determined by $g_*=106.75$.  Here, we should emphasize the difference with the traditional curvaton mechanism, in which curvaton dubs as an independent field leading to the decay of curvaton after the actual preheating. However, our model is different since the curvaton generates and practically disappears as the inflaton decays, to be more precise, the curvaton formed as the inflaton starts to decay, and meanwhile, the curvaton will disappear as the inflaton finishes the decay process due to the coupling between the curvaton and inflaton. During the whole process of preheating for inflaton, the PBH will be formed due to the instability of curvaton. That is the reason we mentioned this process happens during the deep radiation period. 

To investigate the abundance of PBH with mass $M$, we need the mass fraction $\beta(M)$ against the total energy at the formation of PBH. Under the assumption of the distribution is Gaussian, it can be expressed by \cite{Tada:2019amh, Young:2014ana}
\begin{equation}
	\beta(M)=\frac{\rho_{\rm PBH}}{\rho_{\rm total}}=\frac{1}{2}\rm erfc\bigg(\frac{\delta_c}{\sqrt{2\sigma^2(M)}}\bigg),
	\label{mass fraction of pbh}
\end{equation}
where $\rm erfc$ is the complementary error function and $\delta_c$ denotes the threshold of perturbation at the formation of PBH whose value is around $0.4$ in light of \cite{Musco:2012au,Harada:2013epa},  in which Ref. \cite{Martin:2019nuw,Hidalgo:2017dfp} also proposed the value of $\delta_c$, even Ref. \cite{Escriva:2019phb,Escriva:2020tak} discussed the value of $\delta_c$ is also relavant with compaction function leading to a tiny deviation with $0.4$. $\delta(M)$ is the variance of the density perturbation with mass $M$ for PBH which can be associated with power spectrum, 
\begin{equation}
	\sigma^2(M(k))=\int d\ln q W^2(qk^{-1})\frac{16}{81}(qk^{-1})^4 P_\zeta(q),
	\label{variance of pbh}
\end{equation}
where $W(x)=\exp (-x^2/2)$ (Gaussian window function). Once having these two physical quantities, we can define the abundance of PBH, namely the fraction of PBH to the total DM, 
\begin{equation}
	f_{\rm PBH}=\frac{\Omega_{\rm PBH}}{\Omega_{\rm DM}}=2.0\times 10^8\bigg(\frac{\gamma}{0.2}\bigg)^{1/2}\bigg(\frac{10.75}{g_*}\bigg)^{1/4}\bigg(\frac{M_{\odot}}{M}\bigg)^{1/2}\beta(M),
	\label{fm}
\end{equation}
where $\Omega_{\rm DM}$ is the current energy density of DM from Planck 2018 results giving its value is around $\Omega_{\rm DM}h^2 \approx 0.12$. In light of basic estimation for the abundance of PBH, the amount of power spectrum of $P_\zeta$ should reach the order of $10^{-2}$ to produce the sizable PBH abundance on small scales, in which we have given our numerical simulation from figure \ref{full power spectrum}. 

\begin{figure}[h!]
	\centering
	\includegraphics[height=9cm, width=13cm]{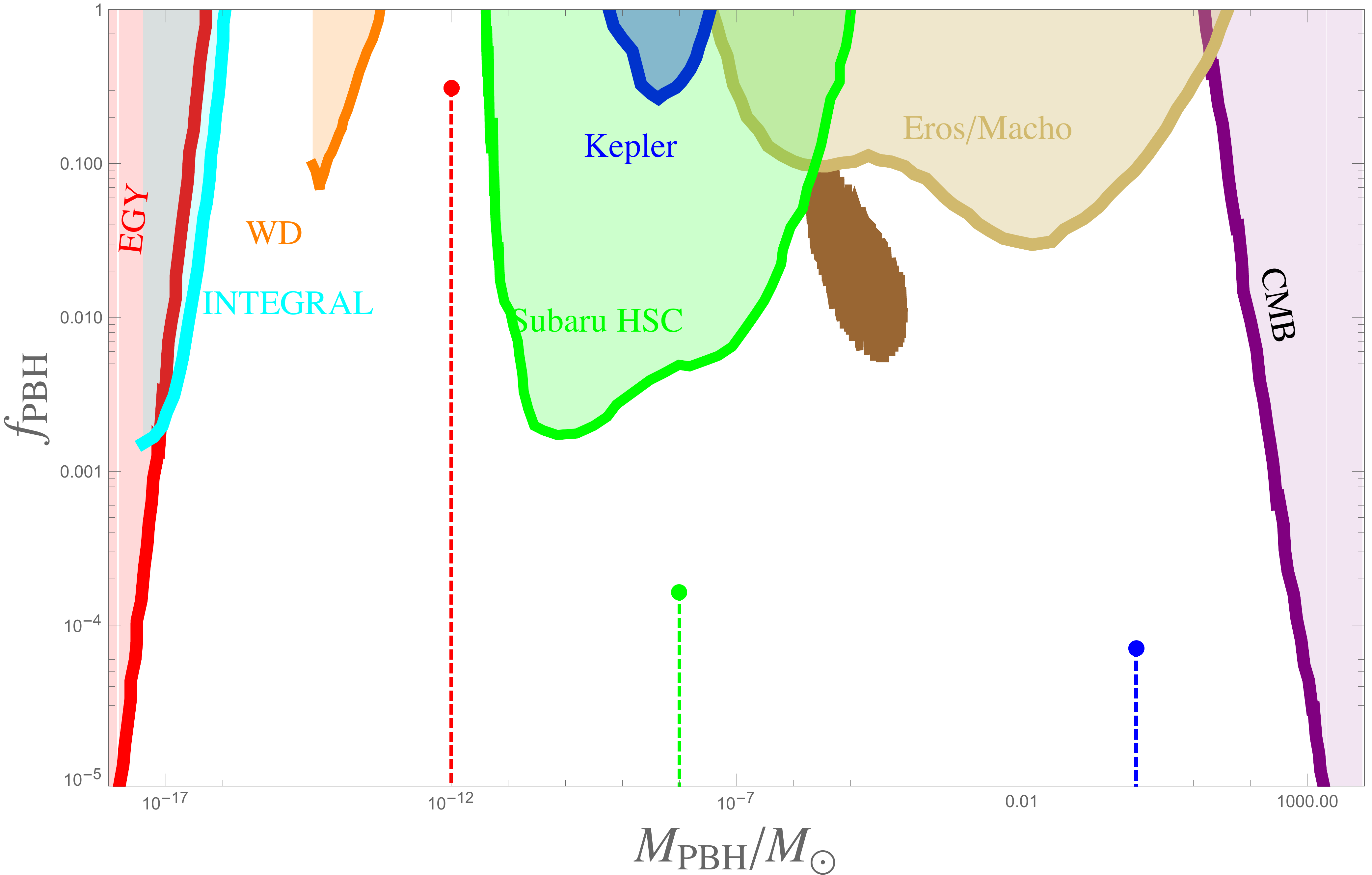}
	\vskip -0.4cm
	\caption{{\it The plot of the abundance of PBH (\ref{fm}):} The horizontal line corresponds to the ratio of $M_{\rm PBH}$ to $M_{\rm \odot}$ whose range is from $10^{-19}$ to $10^4$, which could cover the whole mass range of PBH. The vertical line is the abundance of PBH, in which its corresponding range is from $10^{-5}$ to $1$. The red, green, and blue dashed lines correspond to the cases as depicted in figure \ref{full power spectrum} with various masses of the inflaton. The blue allowed region comes via the ultrashort-timescale microlensing events of OGLE data \cite{Niikura:2019kqi}. The other shadow part denote the current allowed observations, the extragalactic gamma-rays of  PBH evaporation (EG$\gamma$) \cite{Carr:2009jm}, the white dwarfs explosion (WD) \cite{Graham:2015apa}, the galactic center $511~keV$ gamma-ray line (INTEGRAL) \cite{Laha:2019ssq} (Ref. \cite{Dasgupta:2019cae} discussed the allowing maximum rotation can significantly improve and extend the constraints from $511~keV$ in higher mass windows), the microlensing events with Subaru HSC (Subaru HSC) \cite{Acharyya:2019nwy}, with the Kepler satellite (Kepler) \cite{Griest:2013esa}, the EROS/MACHO (EROS/MACHO) \cite{EROS-2:2006ryy}, and the constraints from CMB (CMB) \cite{Ali-Haimoud:2016mbv}. }
	\label{fbh}
\end{figure}

Figure \ref{fbh} indicates the abundance of PBH to the total energy density of DM in terms of the running curvaton scenario. There are three cases corresponding to the various masses as showing in figure \ref{full power spectrum}, in which their mass are $10^{-12} M_\odot $, $10^{-8}M_\odot$ and $10^{0} M_\odot$, respectively, in which their the corresponding mass is obtained by the definition of $\mu_k$ as taking its maximal value that is namely $2k=m$ as setting $m_1$, $m_2$, $m_3$, $\rm e.t.c$. One these three cases, case one (the red point) could result in the main component of DM whose percentage is around $76.5\%$, namely, it could account for the DM in some sense. As for case 2 (green point) and case 3 (blue point), they are far from the observational constraints, especially compared with Kepler \cite{Griest:2013esa} and Subaru HSC \cite{Acharyya:2019nwy}. Case 3 has a similar situation compared with CMB \cite{Ali-Haimoud:2016mbv}.  However, Ref. \cite{Hutsi:2020sol} claimed that the mass range of PBH is from $2$ to $400$ solar mass, which could only account for the $0.2\%$ DM, meanwhile \cite{Carr:2020gox} also discussed that the less contribution of DM will play an important role and supply a significant probe for the early Universe. 

\begin{figure}[h!]
	\centering
	\includegraphics[height=8.9cm, width=8.52cm]{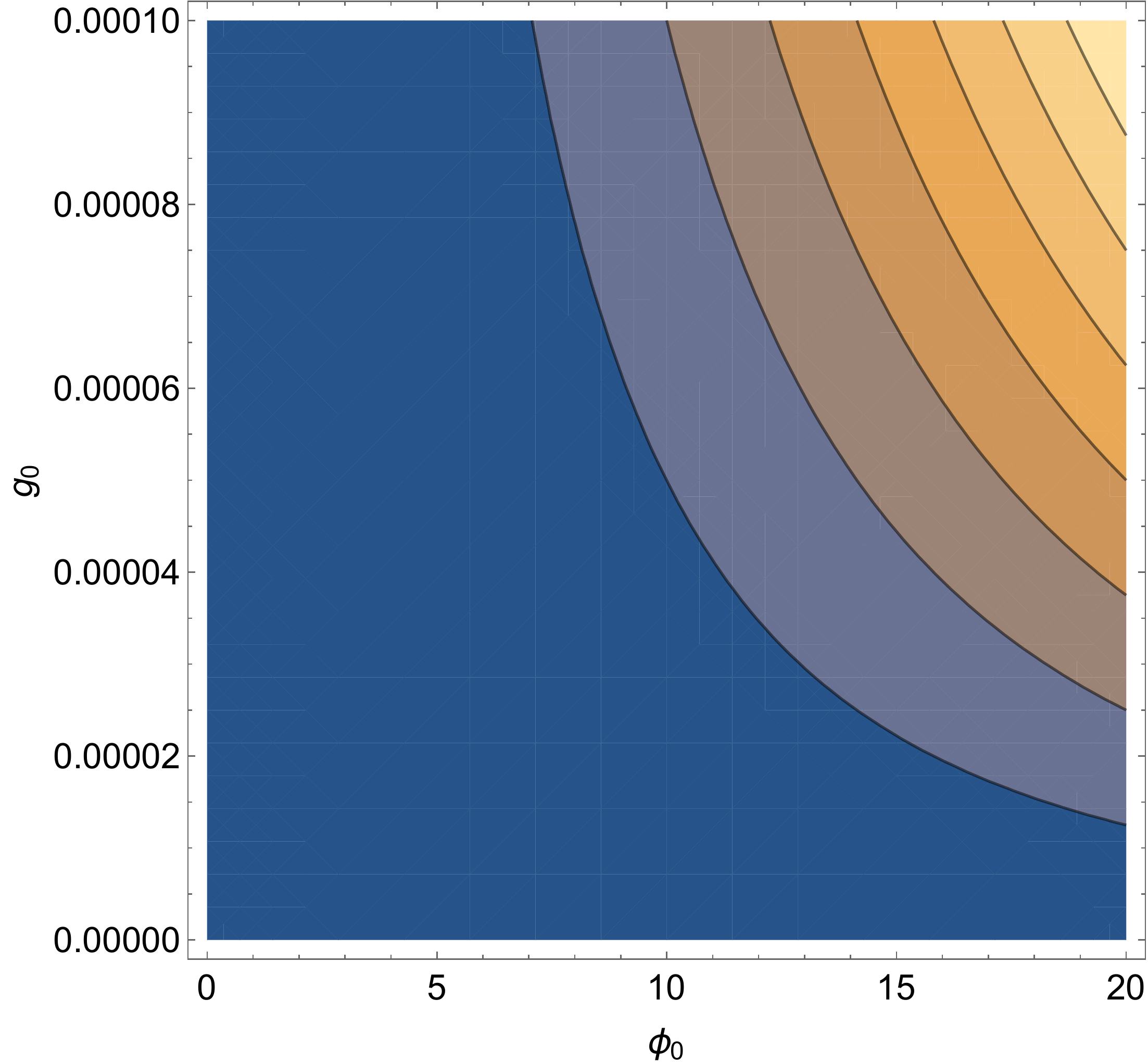}
	\includegraphics[height=8.7cm, width=1.05cm ]{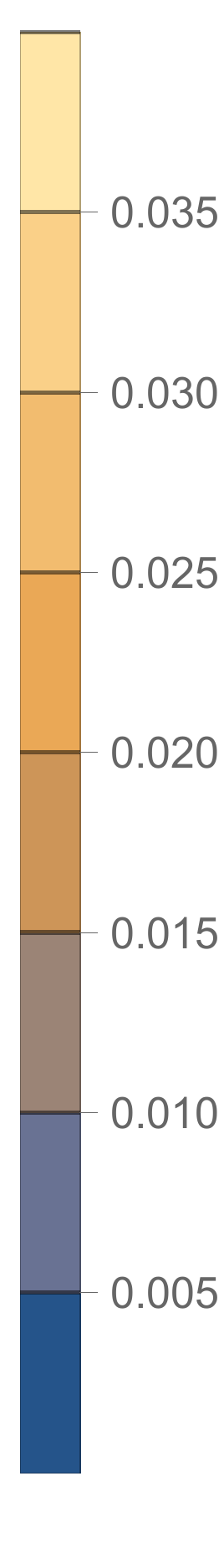}
	\vskip -0.4cm
	\caption{{\it Contour Plot of $q_k$ via eq. (\ref{correspondence1}):} The horizontal line corresponds to the amplitude of inflaton whose range is $0\leqslant phi_0\leqslant20$ determined by the e-folding number is around $60$. The vertical line denotes the value of $g_0$ whose range locates from $0$ to $0.0001$, we have set $M_{\rm P}=1$ and $a=1$.  The right panel shows that the value of $q_k$ matching its corresponding color.}
	\label{qk}
\end{figure}

Here, we only vary with mass of inflaton, in which the potential of inflaton can be constrained as follows \cite{Liu:2018hno},
\begin{equation}
	\frac{V(\phi)}{M_{\rm P}^4}\approx 3.0\times 10^{-10} \bigg(\frac{r_*}{5\times 10^{-3}}\bigg)\bigg(\frac{A_s}{2.1\times 10^{-9}}\bigg),
	\label{constraint of potential}
\end{equation}
where $r_*$ is the tensor-to-scalar ratio whose value is less than $0.06$ and $A_s$ is the amplitude of the power spectrum of curvature perturbation whose value is around $2.1\times 10^{-9}$. The constraint comes from COBE normalization, which tells the range of inflationary potential during inflation. After inflation, the Universe undergoes the preheating period, and the value of inflationary potential is smaller than $3.0\times 10^{-9}$ since the value of the inflaton field will be decreased by transferring the energy into other fundamental particles (\rm i.e. Higgs particles, and other Fermions). Thus, it gives us lots of freedom to simulate the range of value for inflaton mass, which we have shown in figure \ref{fbh} and \ref{full power spectrum}. Consequently, our mechanism will nicely recover the whole mass range of PBH. In some sense, it could account for the origin of DM. In our numerical simulation, the parameters we adapted belong to the narrow resonance, namely $q_k\ll 1$. In figure \ref{qk}, we have plotted the range $q_k$, in which we have set $M_{\rm P}=1$ and $a=1$ (scale factor), and the other parameters can be explicitly shown in this figure. From figure \ref{qk}, it indicates that the range of $g_0$ cannot be large in order to obtain $q_k\ll 1$, whose range is compatible with \cite{Torres-Lomas:2014bua}. As for the broad narrow resonance, it will also lead to the sufficient production of PBH. However, the over-production of PBH is not a generic feature even in broad resonance, namely $q\gg 1$ \cite{Torres-Lomas:2014bua}, which is determined by the coupling constant between the inflaton field and target field. By considering our model, the curvaton comes via the decay of inflaton, thus the formation of PBH is an inevitable process whose value of PBH abundance is sufficient analysis in Ref. \cite{Suyama:2004mz}, in which the criteria of PBH formation is the duration time of preheating, thus the power spectrum is mainly characterized by the scale, not the key feature of instability of Mathieu equation describing the preheating period. To depict this feature, we characterize the power spectrum by the $\delta$ function highly relevant to the scale on small scales. For the large scales, it is almost the Gaussian for power spectrum (nearly scale-invariant).

In this section, we have thoroughly investigated the formation of PBH during the preheating period. The key ingredient for generating the PBH is utilizing the instability of the Mathieu equation characterized by the $\delta$ functions in the power spectrum at small scales. In the next section, we will proceed with the study of the equation of state $EoS$ to investigate the dark energy epoch.

\section{The late time evolution of curvaton field}
\label{late time}

The curvaton mechanism occurs during the preheating period, meanwhile, it comes via the decay of inflaton. From action (\ref{total action}), it indicates that the main contribution will be disappeared as the inflaton finish the decay process. To clarify this situation, we will make a plot of potential consisting of the inflaton and curvaton, however, the curvaton comes via the transferring of energy of inflaton, meanwhile, there are lots of uncertainties about preheating mechanism. We will only show varying values of inflaton and curvaton, in which the inflaton field will be decreased and the curvaton will be enhanced. 

\begin{figure}[h!]
	\centering
	\includegraphics[height=8.9cm, width=8.52cm]{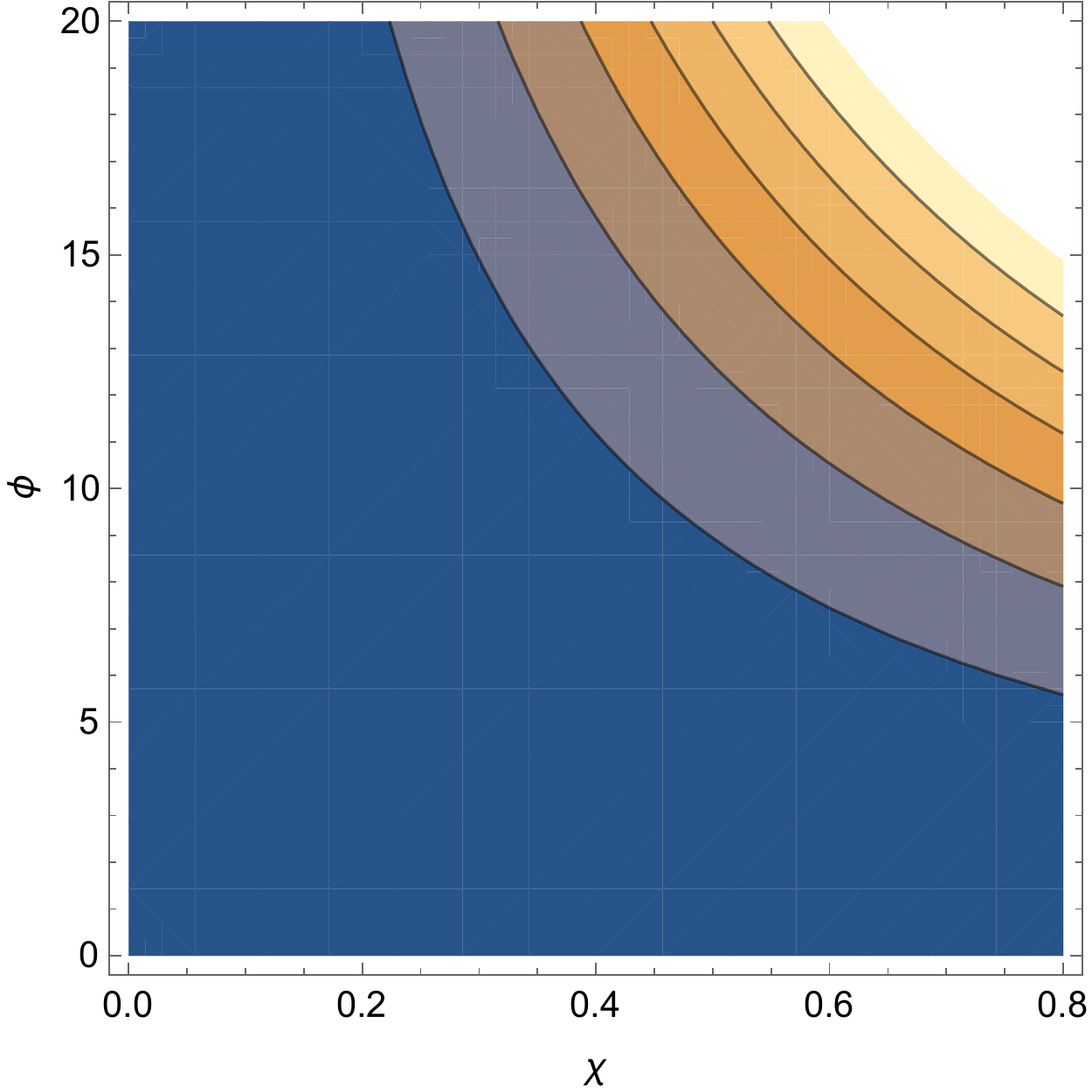}
	\includegraphics[height=8.67cm, width=1.0cm ]{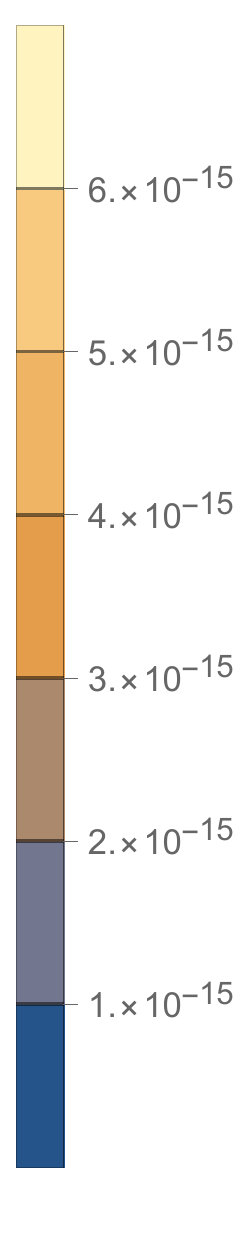}
	\vskip -0.4cm
	\caption{{\it Contour Plot of curvaton.} It shows that the potential of curvaton field, in which $V(\chi)=\frac{g_0\chi^2}{M_{\rm P}}V(\phi)+\lambda_0\exp(-\lambda_1\frac{\chi}{M_{\rm P}})$ with $V(\phi)=\frac{1}{2}m^2\phi^2$ as an instance. we have set $M_{\rm P}=1$, $g_0=0.001$ and $\lambda_1$ has been obsorbed in the curvaton field.} 
	\label{curvaton potential}
\end{figure}
In figure \ref{curvaton potential}, we have shown the potential of curvaton during preheating which just is considered an illustration of the range of curvaton. The field value of curvaton comes via the inflaton decay. Consequently, its contour plot is highly probably different from figure \ref{curvaton potential}, however, it will not change the range of field for curvaton. We could see that the range of curvaton is around $10^{-15}$ in Planck units as taking proper parameters, which means that its energy density will be dominant after inflaton decay but much more tiny compared with the energy scope of inflationary potential (matching with curvaton assumption). As for the curvaton decay, it will nearly disappear as the completion of inflaton decay only keeping the relic of the exponential potential of $curvaton$, namely, it is almost impossible to decay into other particles determined by its effective mass $m_{\rm eff}^2=\frac{d^2V(\chi)}{d\chi^2}$, since Ref. \cite{Liu:2019xhn} has indicated that the order of effective mass for curvaton is of order cosmological constant that means that the effective mass of curvaton is much smaller compared with all kinds of fundamental particles including Higgs particles, fermions, and gauge field particles, $\rm e.t.c.$, in which the new place of our model is that the lifetime of curvaton is almost the same with inflaton since the main mass part of curvaton is proportional to the inflationary potential. Simple and explicit analysis for the curvaton potential is taken into account, in which it will be naturally approached to a constant being of the order of cosmological constant dubbed as a role of dark energy. 

From this section, we give a simple analysis that shows that the range of curvaton is less than $10^{-15}$ in Planck units. In this scope, the potential of curvaton is dominant during preheating but much more tiny compared with the inflationary energy scale matching the assumption of curvaton. At late times, the curvaton field will also decay into other fundamental particles. However, there is a relic of the exponential potential of curvaton playing a role in cosmological constant, which can be dubbed as the dark energy dominating the current epoch. In some sense, this curvaton mechanism adding the formation of PBH during preheating could unify the dark energy and dark matter. 

\section{Conclusion and discussions}
\label{conclusion}
In this paper, we have investigated the formation of PBH during the preheating period in light of the instability of the Mathieu equation. Being different from previous work \cite{Martin:2019nuw,Torres-Lomas:2014bua,Suyama:2004mz}, we will implement the running mass of curvaton which is proportional to \eqref{approximating potential} to investigate the formation of PBH during the deeply preheating period (induced by the second inflation). Thus, the mass of curvaton will be varying from COBE normalization to the DE scale determined by $V_{\rm eff}=\frac{d^2V}{d\chi^2}$ as the inflaton finishes the decay. Due to its huge range, we can relate the exponential growth of curvature perturbation to the mass scale which is the instability band corresponding to the $\tilde{k}$. Then, we can adapt the property in light of the property of $\delta$ functions, the other energy scale is almost scale-invariant as shown in eq. (\ref{the whole power spctrum}). Thus, the simple analytical power spectrum perfectly agreed with observations whose detail can be seen in figure \ref{full power spectrum}.

Once obtained this key result of this paper, we can numerically simulate the abundance of PBH among the DM. Figure \ref{fbh} clearly indicates the value of $f_{\rm PBH}$ with some specific values of mass corresponding to the certain $\tilde{k}$.  Here, we only use the property of the Lyapunov index to find the corresponding various values of $k$ could simulate different values of $f_{\rm PBH}$ as choosing proper parameters. Especially for the case $1$ of figure \ref{fbh}, it shows that the value of abundance of PBH could reach the $75\%$ of DM which could account for the DM in some sense.  During the preheating period, we have shown that the potential is of an order of $10^{-15}$ dominating the main content of the Universe and meanwhile being consistent with the assumption of curvaton. At late times, as the inflaton almost completes the process of decay (curvaton disappears), there is a relic of curvaton exponential potential that will be dubbed as the role of dark energy as illustrated in section \ref{late time}. Thus, our model could unify dark energy and dark matter.

Our model is highly relevant to the coupling structure. Consequently, we could use this similar mechanism to explore the possibility of the formation of the Higgs field during preheating, and the mass of the Higgs field is stringently constrained by the observation and also its rich decay channels. It sheds a light on exploring Higgs physics. Finally, we should emphasize that our non-Guaassianity parameter can be extended into a new one associated with $w$ (EoS) and $r_{\rm decay}$ \cite{Liu:2020zlr}.

\section*{Acknowledgements}
LH is funded by the Hunan Provincial Department of Education, NO. 19B464, and the National Natural Science Foundation of China Grants (NSFC) NO. 12165009. We are grateful for the discussions with Diego Cruces for the value of $\delta$ and the numerical simulation of Wu-Long Xu.


\section*{References}
\begin{thebibliography}{99}
	
	
	
	
	
	\bibitem{Liu:2019xhn}
	L.~H.~Liu and W.~L.~Xu,
	Chin. Phys. C \textbf{44}, no.8, 085103 (2020)
	doi:10.1088/1674-1137/44/8/085103
	[arXiv:1911.10542 [astro-ph.CO]].
	
	
	\bibitem{Hawking:1971ei}
	S.~Hawking,
	Mon. Not. Roy. Astron. Soc. \textbf{152}, 75 (1971)
	
	\bibitem{Carr:1974nx}
	B.~J.~Carr and S.~W.~Hawking,
	Mon. Not. Roy. Astron. Soc. \textbf{168}, 399-415 (1974)
	
	\bibitem{Grindlay:1975eb}
	J.~E.~Grindlay, H.~F.~Helmken, R.~H.~Brown, J.~Davis and L.~R.~Allen,
	Astrophys. J. \textbf{201}, 82-89 (1975)
	doi:10.1086/153861
	

\bibitem{Ivanov:1994pa}
P.~Ivanov, P.~Naselsky and I.~Novikov,
Phys. Rev. D \textbf{50}, 7173-7178 (1994)
doi:10.1103/PhysRevD.50.7173


\bibitem{Carr:2016drx}
B.~Carr, F.~Kuhnel and M.~Sandstad,
Phys. Rev. D \textbf{94}, no.8, 083504 (2016)
doi:10.1103/PhysRevD.94.083504
[arXiv:1607.06077 [astro-ph.CO]].


\bibitem{Gaggero:2016dpq}
D.~Gaggero, G.~Bertone, F.~Calore, R.~M.~T.~Connors, M.~Lovell, S.~Markoff and E.~Storm,
Phys. Rev. Lett. \textbf{118}, no.24, 241101 (2017)
doi:10.1103/PhysRevLett.118.241101
[arXiv:1612.00457 [astro-ph.HE]].


\bibitem{Inomata:2017okj}
K.~Inomata, M.~Kawasaki, K.~Mukaida, Y.~Tada and T.~T.~Yanagida,
Phys. Rev. D \textbf{96}, no.4, 043504 (2017)
doi:10.1103/PhysRevD.96.043504
[arXiv:1701.02544 [astro-ph.CO]].



\bibitem{Kovetz:2017rvv}
E.~D.~Kovetz,
Phys. Rev. Lett. \textbf{119}, no.13, 131301 (2017)
doi:10.1103/PhysRevLett.119.131301
[arXiv:1705.09182 [astro-ph.CO]].

\bibitem{Georg:2017mqk}
J.~Georg and S.~Watson,
JHEP \textbf{09}, 138 (2017)
doi:10.1007/JHEP09(2017)138
[arXiv:1703.04825 [astro-ph.CO]].


\bibitem{LIGOScientific:2016aoc}
B.~P.~Abbott \textit{et al.} [LIGO Scientific and Virgo],
Phys. Rev. Lett. \textbf{116}, no.6, 061102 (2016)
doi:10.1103/PhysRevLett.116.061102
[arXiv:1602.03837 [gr-qc]].


\bibitem{LIGOScientific:2016emj}
B.~P.~Abbott \textit{et al.} [LIGO Scientific and Virgo],
Phys. Rev. Lett. \textbf{116}, no.13, 131103 (2016)
doi:10.1103/PhysRevLett.116.131103
[arXiv:1602.03838 [gr-qc]].


\bibitem{LIGOScientific:2017bnn}
B.~P.~Abbott \textit{et al.} [LIGO Scientific and VIRGO],
Phys. Rev. Lett. \textbf{118}, no.22, 221101 (2017)
[erratum: Phys. Rev. Lett. \textbf{121}, no.12, 129901 (2018)]
doi:10.1103/PhysRevLett.118.221101
[arXiv:1706.01812 [gr-qc]].

\bibitem{Bird:2016dcv}
S.~Bird, I.~Cholis, J.~B.~Mu\~noz, Y.~Ali-Ha\"\i{}moud, M.~Kamionkowski, E.~D.~Kovetz, A.~Raccanelli and A.~G.~Riess,
Phys. Rev. Lett. \textbf{116}, no.20, 201301 (2016)
doi:10.1103/PhysRevLett.116.201301
[arXiv:1603.00464 [astro-ph.CO]].


\bibitem{Sasaki:2016jop}
M.~Sasaki, T.~Suyama, T.~Tanaka and S.~Yokoyama,
Phys. Rev. Lett. \textbf{117}, no.6, 061101 (2016)
[erratum: Phys. Rev. Lett. \textbf{121}, no.5, 059901 (2018)]
doi:10.1103/PhysRevLett.117.061101
[arXiv:1603.08338 [astro-ph.CO]].



\bibitem{Garcia-Bellido:1996mdl}
J.~Garcia-Bellido, A.~D.~Linde and D.~Wands,
Phys. Rev. D \textbf{54}, 6040-6058 (1996)
doi:10.1103/PhysRevD.54.6040
[arXiv:astro-ph/9605094 [astro-ph]].


\bibitem{Garcia-Bellido:2017mdw}
J.~Garcia-Bellido and E.~Ruiz Morales,
Phys. Dark Univ. \textbf{18}, 47-54 (2017)
doi:10.1016/j.dark.2017.09.007
[arXiv:1702.03901 [astro-ph.CO]].

\bibitem{Domcke:2017fix}
V.~Domcke, F.~Muia, M.~Pieroni and L.~T.~Witkowski,
JCAP \textbf{07}, 048 (2017)
doi:10.1088/1475-7516/2017/07/048
[arXiv:1704.03464 [astro-ph.CO]].



\bibitem{Kannike:2017bxn}
K.~Kannike, L.~Marzola, M.~Raidal and H.~Veerm\"ae,
JCAP \textbf{09}, 020 (2017)
doi:10.1088/1475-7516/2017/09/020
[arXiv:1705.06225 [astro-ph.CO]].


\bibitem{Carr:2017edp}
B.~Carr, T.~Tenkanen and V.~Vaskonen,
Phys. Rev. D \textbf{96}, no.6, 063507 (2017)
doi:10.1103/PhysRevD.96.063507
[arXiv:1706.03746 [astro-ph.CO]].


\bibitem{Ballesteros:2017fsr}
G.~Ballesteros and M.~Taoso,
Phys. Rev. D \textbf{97}, no.2, 023501 (2018)
doi:10.1103/PhysRevD.97.023501
[arXiv:1709.05565 [hep-ph]].

\bibitem{Hertzberg:2017dkh}
M.~P.~Hertzberg and M.~Yamada,
Phys. Rev. D \textbf{97}, no.8, 083509 (2018)
doi:10.1103/PhysRevD.97.083509
[arXiv:1712.09750 [astro-ph.CO]].

\bibitem{Franciolini:2018vbk}
G.~Franciolini, A.~Kehagias, S.~Matarrese and A.~Riotto,
JCAP \textbf{03}, 016 (2018)
doi:10.1088/1475-7516/2018/03/016
[arXiv:1801.09415 [astro-ph.CO]].

\bibitem{Kohri:2018qtx}
K.~Kohri and T.~Terada,
Class. Quant. Grav. \textbf{35}, no.23, 235017 (2018)
doi:10.1088/1361-6382/aaea18
[arXiv:1802.06785 [astro-ph.CO]].

\bibitem{Ozsoy:2018flq}
O.~\"Ozsoy, S.~Parameswaran, G.~Tasinato and I.~Zavala,
JCAP \textbf{07}, 005 (2018)
doi:10.1088/1475-7516/2018/07/005
[arXiv:1803.07626 [hep-th]].


\bibitem{Biagetti:2018pjj}
M.~Biagetti, G.~Franciolini, A.~Kehagias and A.~Riotto,
JCAP \textbf{07}, 032 (2018)
doi:10.1088/1475-7516/2018/07/032
[arXiv:1804.07124 [astro-ph.CO]].



\bibitem{Belotsky:2014kca}
K.~M.~Belotsky, A.~D.~Dmitriev, E.~A.~Esipova, V.~A.~Gani, A.~V.~Grobov, M.~Y.~Khlopov, A.~A.~Kirillov, S.~G.~Rubin and I.~V.~Svadkovsky,
Mod. Phys. Lett. A \textbf{29}, no.37, 1440005 (2014)
doi:10.1142/S0217732314400057
[arXiv:1410.0203 [astro-ph.CO]].


\bibitem{Khlopov:2008qy}
M.~Y.~Khlopov,
Res. Astron. Astrophys. \textbf{10}, 495-528 (2010)
doi:10.1088/1674-4527/10/6/001
[arXiv:0801.0116 [astro-ph]].



\bibitem{Martin:2012pe}
J.~Martin, H.~Motohashi and T.~Suyama,
Phys. Rev. D \textbf{87}, no.2, 023514 (2013)
doi:10.1103/PhysRevD.87.023514
[arXiv:1211.0083 [astro-ph.CO]].


\bibitem{Motohashi:2014ppa}
H.~Motohashi, A.~A.~Starobinsky and J.~Yokoyama,
JCAP \textbf{09}, 018 (2015)
doi:10.1088/1475-7516/2015/09/018
[arXiv:1411.5021 [astro-ph.CO]].


\bibitem{Germani:2017bcs}
C.~Germani and T.~Prokopec,
Phys. Dark Univ. \textbf{18}, 6-10 (2017)
doi:10.1016/j.dark.2017.09.001
[arXiv:1706.04226 [astro-ph.CO]].


\bibitem{Motohashi:2017kbs}
H.~Motohashi and W.~Hu,
Phys. Rev. D \textbf{96}, no.6, 063503 (2017)
doi:10.1103/PhysRevD.96.063503
[arXiv:1706.06784 [astro-ph.CO]].


\bibitem{Ezquiaga:2017fvi}
J.~M.~Ezquiaga, J.~Garcia-Bellido and E.~Ruiz Morales,
Phys. Lett. B \textbf{776}, 345-349 (2018)
doi:10.1016/j.physletb.2017.11.039
[arXiv:1705.04861 [astro-ph.CO]].

\bibitem{Ballesteros:2018wlw}
G.~Ballesteros, J.~Beltran Jimenez and M.~Pieroni,
JCAP \textbf{06}, 016 (2019)
doi:10.1088/1475-7516/2019/06/016
[arXiv:1811.03065 [astro-ph.CO]].



\bibitem{Passaglia:2018ixg}
S.~Passaglia, W.~Hu and H.~Motohashi,
Phys. Rev. D \textbf{99}, no.4, 043536 (2019)
doi:10.1103/PhysRevD.99.043536
[arXiv:1812.08243 [astro-ph.CO]].


\bibitem{Sasaki:2018dmp}
M.~Sasaki, T.~Suyama, T.~Tanaka and S.~Yokoyama,
Class. Quant. Grav. \textbf{35}, no.6, 063001 (2018)
doi:10.1088/1361-6382/aaa7b4
[arXiv:1801.05235 [astro-ph.CO]].



\bibitem{Fu:2019ttf}
C.~Fu, P.~Wu and H.~Yu,
Phys. Rev. D \textbf{100}, no.6, 063532 (2019)
doi:10.1103/PhysRevD.100.063532
[arXiv:1907.05042 [astro-ph.CO]].


\bibitem{Fu:2019vqc}
C.~Fu, P.~Wu and H.~Yu,
Phys. Rev. D \textbf{101}, no.2, 023529 (2020)
doi:10.1103/PhysRevD.101.023529
[arXiv:1912.05927 [astro-ph.CO]].


\bibitem{Dalianis:2019vit}
I.~Dalianis, S.~Karydas and E.~Papantonopoulos,
JCAP \textbf{06}, 040 (2020)
doi:10.1088/1475-7516/2020/06/040
[arXiv:1910.00622 [astro-ph.CO]].


\bibitem{Lin:2020goi}
J.~Lin, Q.~Gao, Y.~Gong, Y.~Lu, C.~Zhang and F.~Zhang,
Phys. Rev. D \textbf{101}, no.10, 103515 (2020)
doi:10.1103/PhysRevD.101.103515
[arXiv:2001.05909 [gr-qc]].

\bibitem{Braglia:2020eai}
M.~Braglia, D.~K.~Hazra, F.~Finelli, G.~F.~Smoot, L.~Sriramkumar and A.~A.~Starobinsky,
JCAP \textbf{08}, 001 (2020)
doi:10.1088/1475-7516/2020/08/001
[arXiv:2005.02895 [astro-ph.CO]].


\bibitem{Gundhi:2020zvb}
A.~Gundhi and C.~F.~Steinwachs,
Eur. Phys. J. C \textbf{81}, no.5, 460 (2021)
doi:10.1140/epjc/s10052-021-09225-2
[arXiv:2011.09485 [hep-th]].

\bibitem{Cheong:2019vzl}
D.~Y.~Cheong, S.~M.~Lee and S.~C.~Park,
JCAP \textbf{01}, 032 (2021)
doi:10.1088/1475-7516/2021/01/032
[arXiv:1912.12032 [hep-ph]].




\bibitem{Armendariz-Picon:1999hyi}
C.~Armendariz-Picon, T.~Damour and V.~F.~Mukhanov,
Phys. Lett. B \textbf{458}, 209-218 (1999)
doi:10.1016/S0370-2693(99)00603-6
[arXiv:hep-th/9904075 [hep-th]].



\bibitem{Garriga:1999vw}
J.~Garriga and V.~F.~Mukhanov,
Phys. Lett. B \textbf{458}, 219-225 (1999)
doi:10.1016/S0370-2693(99)00602-4
[arXiv:hep-th/9904176 [hep-th]].



\bibitem{Kobayashi:2010cm}
T.~Kobayashi, M.~Yamaguchi and J.~Yokoyama,
Phys. Rev. Lett. \textbf{105}, 231302 (2010)
doi:10.1103/PhysRevLett.105.231302
[arXiv:1008.0603 [hep-th]].


\bibitem{Kobayashi:2011nu}
T.~Kobayashi, M.~Yamaguchi and J.~Yokoyama,
Prog. Theor. Phys. \textbf{126}, 511-529 (2011)
doi:10.1143/PTP.126.511
[arXiv:1105.5723 [hep-th]].


\bibitem{Cai:2018tuh}
Y.~F.~Cai, X.~Tong, D.~G.~Wang and S.~F.~Yan,
Phys. Rev. Lett. \textbf{121}, no.8, 081306 (2018)
doi:10.1103/PhysRevLett.121.081306
[arXiv:1805.03639 [astro-ph.CO]].


\bibitem{Cai:2019jah}
Y.~F.~Cai, C.~Chen, X.~Tong, D.~G.~Wang and S.~F.~Yan,
Phys. Rev. D \textbf{100}, no.4, 043518 (2019)
doi:10.1103/PhysRevD.100.043518
[arXiv:1902.08187 [astro-ph.CO]].



\bibitem{Chen:2020uhe}
C.~Chen, X.~H.~Ma and Y.~F.~Cai,
Phys. Rev. D \textbf{102}, no.6, 063526 (2020)
doi:10.1103/PhysRevD.102.063526
[arXiv:2003.03821 [astro-ph.CO]].

\bibitem{Chen:2019zza}
C.~Chen and Y.~F.~Cai,
JCAP \textbf{10}, 068 (2019)
doi:10.1088/1475-7516/2019/10/068
[arXiv:1908.03942 [astro-ph.CO]].






\bibitem{Kamenshchik:2018sig}
A.~Y.~Kamenshchik, A.~Tronconi, T.~Vardanyan and G.~Venturi,
Phys. Lett. B \textbf{791}, 201-205 (2019)
doi:10.1016/j.physletb.2019.02.036
[arXiv:1812.02547 [gr-qc]].




\bibitem{Cai:2019bmk}
R.~G.~Cai, Z.~K.~Guo, J.~Liu, L.~Liu and X.~Y.~Yang,
JCAP \textbf{06}, 013 (2020)
doi:10.1088/1475-7516/2020/06/013
[arXiv:1912.10437 [astro-ph.CO]].



\bibitem{Wang:2021kbh}
Q.~Wang, Y.~C.~Liu, B.~Y.~Su and N.~Li,
Phys. Rev. D \textbf{104}, no.8, 083546 (2021)
doi:10.1103/PhysRevD.104.083546
[arXiv:2111.10028 [astro-ph.CO]].


\bibitem{Enqvist:2001zp}
K.~Enqvist and M.~S.~Sloth,
``Adiabatic CMB perturbations in pre - big bang string cosmology,''
Nucl.\ Phys.\ B {\bf 626} (2002) 395
doi:10.1016/S0550-3213(02)00043-3
[hep-ph/0109214].

\bibitem{Lyth:2001nq}
D.~H.~Lyth and D.~Wands,
``Generating the curvature perturbation without an inflaton,''
Phys.\ Lett.\ B {\bf 524} (2002) 5
doi:10.1016/S0370-2693(01)01366-1
[hep-ph/0110002].

\bibitem{Moroi:2001ct}
T.~Moroi and T.~Takahashi,
``Effects of cosmological moduli fields on cosmic microwave background,''
Phys.\ Lett.\ B {\bf 522} (2001) 215
Erratum: [Phys.\ Lett.\ B {\bf 539} (2002) 303]
doi:10.1016/S0370-2693(02)02070-1, 10.1016/S0370-2693(01)01295-3
[hep-ph/0110096].


\bibitem{Gong:2016yyb}
J.~O.~Gong, N.~Kitajima and T.~Terada,
JCAP {\bf 1703} (2017) 053
doi:10.1088/1475-7516/2017/03/053
[arXiv:1611.08975 [hep-ph]].


\bibitem{Kawasaki:2012wr}
M.~Kawasaki, N.~Kitajima and T.~T.~Yanagida,
Phys.\ Rev.\ D {\bf 87} (2013) no.6,  063519
doi:10.1103/PhysRevD.87.063519
[arXiv:1207.2550 [hep-ph]].


\bibitem{Ando:2018nge}
K.~Ando, M.~Kawasaki and H.~Nakatsuka,
Phys.\ Rev.\ D {\bf 98} (2018) no.8,  083508
doi:10.1103/PhysRevD.98.083508
[arXiv:1805.07757 [astro-ph.CO]].


\bibitem{Traschen:1990sw}
J.~H.~Traschen and R.~H.~Brandenberger,
``Particle Production During Out-of-equilibrium Phase Transitions,''
Phys.\ Rev.\ D {\bf 42} (1990) 2491.
doi:10.1103/PhysRevD.42.2491

\bibitem{Kofman:1994rk}
L.~Kofman, A.~D.~Linde and A.~A.~Starobinsky,
``Reheating after inflation,''
Phys.\ Rev.\ Lett.\  {\bf 73} (1994) 3195
doi:10.1103/PhysRevLett.73.3195
[hep-th/9405187].

\bibitem{Shtanov:1994ce}
Y.~Shtanov, J.~H.~Traschen and R.~H.~Brandenberger,
``Universe reheating after inflation,''
Phys.\ Rev.\ D {\bf 51} (1995) 5438
doi:10.1103/PhysRevD.51.5438
[hep-ph/9407247].


\bibitem{Prokopec:1996rr}
T.~Prokopec and T.~G.~Roos,
``Lattice study of classical inflaton decay,''
Phys.\ Rev.\ D {\bf 55} (1997) 3768
doi:10.1103/PhysRevD.55.3768
[hep-ph/9610400].

\bibitem{Greene:1997ge}
B.~R.~Greene, T.~Prokopec and T.~G.~Roos,
``Inflaton decay and heavy particle production with negative coupling,''
Phys.\ Rev.\ D {\bf 56} (1997) 6484
doi:10.1103/PhysRevD.56.6484
[hep-ph/9705357].


\bibitem{Kofman:1997yn}
L.~Kofman, A.~D.~Linde and A.~A.~Starobinsky,
``Towards the theory of reheating after inflation,''
Phys.\ Rev.\ D {\bf 56} (1997) 3258
doi:10.1103/PhysRevD.56.3258
[hep-ph/9704452].

\bibitem{Greene:1997fu}
P.~B.~Greene, L.~Kofman, A.~D.~Linde and A.~A.~Starobinsky,
``Structure of resonance in preheating after inflation,''
Phys.\ Rev.\ D {\bf 56} (1997) 6175
doi:10.1103/PhysRevD.56.6175
[hep-ph/9705347].


\bibitem{Liu:2020zzv}
L.~H.~Liu and T.~Prokopec,
JCAP \textbf{06}, 033 (2021)
doi:10.1088/1475-7516/2021/06/033
[arXiv:2005.11069 [astro-ph.CO]].



\bibitem{Cai:2021wzd}
R.~G.~Cai, C.~Chen and C.~Fu,
Phys. Rev. D \textbf{104}, no.8, 083537 (2021)
doi:10.1103/PhysRevD.104.083537
[arXiv:2108.03422 [astro-ph.CO]].



\bibitem{Carrion:2021yeh}
K.~Carrion, J.~C.~Hidalgo, A.~Montiel and L.~E.~Padilla,
JCAP \textbf{07}, 001 (2021)
doi:10.1088/1475-7516/2021/07/001
[arXiv:2101.02156 [astro-ph.CO]].

\bibitem{Moroi:2005np}
T.~Moroi and T.~Takahashi,
Phys. Rev. D \textbf{72}, 023505 (2005)
doi:10.1103/PhysRevD.72.023505
[arXiv:astro-ph/0505339 [astro-ph]].































\bibitem{Akrami:2018odb}
  Y.~Akrami {\it et al.} [Planck Collaboration],
  ``Planck 2018 results. X. Constraints on inflation,''
  arXiv:1807.06211 [astro-ph.CO].



%






%































\bibitem{Baumann:2009ds}
D.~Baumann,
doi:10.1142/9789814327183\_0010
[arXiv:0907.5424 [hep-th]].






























\bibitem{Carr:1975qj}
B.~J.~Carr,
Astrophys. J. \textbf{201}, 1-19 (1975)
doi:10.1086/153853




\bibitem{Tada:2019amh}
Y.~Tada and S.~Yokoyama,
Phys. Rev. D \textbf{100}, no.2, 023537 (2019)
doi:10.1103/PhysRevD.100.023537
[arXiv:1904.10298 [astro-ph.CO]].

\bibitem{Young:2014ana}
S.~Young, C.~T.~Byrnes and M.~Sasaki,
JCAP \textbf{07}, 045 (2014)
doi:10.1088/1475-7516/2014/07/045
[arXiv:1405.7023 [gr-qc]].


\bibitem{Musco:2012au}
I.~Musco and J.~C.~Miller,
Class. Quant. Grav. \textbf{30}, 145009 (2013)
doi:10.1088/0264-9381/30/14/145009
[arXiv:1201.2379 [gr-qc]].



\bibitem{Harada:2013epa}
T.~Harada, C.~M.~Yoo and K.~Kohri,
Phys. Rev. D \textbf{88}, no.8, 084051 (2013)
[erratum: Phys. Rev. D \textbf{89}, no.2, 029903 (2014)]
doi:10.1103/PhysRevD.88.084051
[arXiv:1309.4201 [astro-ph.CO]].


\bibitem{Martin:2019nuw}
J.~Martin, T.~Papanikolaou and V.~Vennin,
JCAP \textbf{01}, 024 (2020)
doi:10.1088/1475-7516/2020/01/024
[arXiv:1907.04236 [astro-ph.CO]].

\bibitem{Hidalgo:2017dfp}
J.~C.~Hidalgo, J.~De Santiago, G.~German, N.~Barbosa-Cendejas and W.~Ruiz-Luna,
Phys. Rev. D \textbf{96}, no.6, 063504 (2017)
doi:10.1103/PhysRevD.96.063504
[arXiv:1705.02308 [astro-ph.CO]].


\bibitem{Escriva:2019phb}
A.~Escriv\`a, C.~Germani and R.~K.~Sheth,
Phys. Rev. D \textbf{101}, no.4, 044022 (2020)
doi:10.1103/PhysRevD.101.044022
[arXiv:1907.13311 [gr-qc]].

\bibitem{Escriva:2020tak}
A.~Escriv\`a, C.~Germani and R.~K.~Sheth,
JCAP \textbf{01}, 030 (2021)
doi:10.1088/1475-7516/2021/01/030
[arXiv:2007.05564 [gr-qc]].



\bibitem{Niikura:2019kqi}
H.~Niikura, M.~Takada, S.~Yokoyama, T.~Sumi and S.~Masaki,
Phys. Rev. D \textbf{99}, no.8, 083503 (2019)
doi:10.1103/PhysRevD.99.083503
[arXiv:1901.07120 [astro-ph.CO]].




\bibitem{Carr:2009jm}
B.~J.~Carr, K.~Kohri, Y.~Sendouda and J.~Yokoyama,
Phys. Rev. D \textbf{81}, 104019 (2010)
doi:10.1103/PhysRevD.81.104019
[arXiv:0912.5297 [astro-ph.CO]].



\bibitem{Graham:2015apa}
P.~W.~Graham, S.~Rajendran and J.~Varela,

novae,''
Phys. Rev. D \textbf{92}, no.6, 063007 (2015)
doi:10.1103/PhysRevD.92.063007
[arXiv:1505.04444 [hep-ph]].


\bibitem{Laha:2019ssq}
R.~Laha,
Phys. Rev. Lett. \textbf{123}, no.25, 251101 (2019)
doi:10.1103/PhysRevLett.123.251101
[arXiv:1906.09994 [astro-ph.HE]].

\bibitem{Dasgupta:2019cae}
B.~Dasgupta, R.~Laha and A.~Ray,
Phys. Rev. Lett. \textbf{125}, no.10, 101101 (2020)
doi:10.1103/PhysRevLett.125.101101
[arXiv:1912.01014 [hep-ph]].



\bibitem{Acharyya:2019nwy}
A.~Acharyya, I.~Agudo, E.~O.~Ang\"uner, R.~Alfaro, J.~Alfaro, C.~Alispach, R.~Aloisio, R.~Alves Batista, J.~P.~Amans and L.~Amati, \textit{et al.}
Astropart. Phys. \textbf{111}, 35-53 (2019)
doi:10.1016/j.astropartphys.2019.04.001
[arXiv:1904.01426 [astro-ph.IM]].



\bibitem{Griest:2013esa}
K.~Griest, A.~M.~Cieplak and M.~J.~Lehner,
Phys. Rev. Lett. \textbf{111}, no.18, 181302 (2013)
doi:10.1103/PhysRevLett.111.181302

\bibitem{EROS-2:2006ryy}
P.~Tisserand \textit{et al.} [EROS-2],
Astron. Astrophys. \textbf{469}, 387-404 (2007)
doi:10.1051/0004-6361:20066017
[arXiv:astro-ph/0607207 [astro-ph]].

\bibitem{Ali-Haimoud:2016mbv}
Y.~Ali-Ha\"\i{}moud and M.~Kamionkowski,
Phys. Rev. D \textbf{95}, no.4, 043534 (2017)
doi:10.1103/PhysRevD.95.043534
[arXiv:1612.05644 [astro-ph.CO]].


\bibitem{Hutsi:2020sol}
G.~H\"utsi, M.~Raidal, V.~Vaskonen and H.~Veerm\"ae,
JCAP \textbf{03}, 068 (2021)
doi:10.1088/1475-7516/2021/03/068
[arXiv:2012.02786 [astro-ph.CO]].


\bibitem{Carr:2020gox}
B.~Carr, K.~Kohri, Y.~Sendouda and J.~Yokoyama,
[arXiv:2002.12778 [astro-ph.CO]].


\bibitem{Liu:2018hno}
L.~H.~Liu, T.~Prokopec and A.~A.~Starobinsky,
Phys. Rev. D \textbf{98}, no.4, 043505 (2018)
doi:10.1103/PhysRevD.98.043505
[arXiv:1806.05407 [gr-qc]].



\bibitem{Torres-Lomas:2014bua}
E.~Torres-Lomas, J.~C.~Hidalgo, K.~A.~Malik and L.~A.~Ure\~na-L\'opez,
Phys. Rev. D \textbf{89}, no.8, 083008 (2014)
doi:10.1103/PhysRevD.89.083008
[arXiv:1401.6960 [astro-ph.CO]].





\bibitem{Suyama:2004mz}
T.~Suyama, T.~Tanaka, B.~Bassett and H.~Kudoh,
Phys. Rev. D \textbf{71}, 063507 (2005)
doi:10.1103/PhysRevD.71.063507
[arXiv:hep-ph/0410247 [hep-ph]].







\bibitem{Liu:2020zlr}
L.~H.~Liu, B.~Liang, Y.~C.~Zhou, X.~D.~Liu, W.~L.~Xu and A.~C.~Li,
Phys. Rev. D \textbf{103}, no.6, 063515 (2021)
doi:10.1103/PhysRevD.103.063515
[arXiv:2007.08278 [astro-ph.CO]].



\end{thebibliography}
\end{document}